\documentclass[twocolumn]{revtex4}
\usepackage{amsmath}
\usepackage{amsfonts}
\usepackage{amssymb}
\usepackage{graphicx}
\usepackage{placeins}
\usepackage{float}
\usepackage{subfigure}
\usepackage{subfigure}
\input{Qcircuit}
\begin{document}
\title{Non-classical Correlations in the Quantum Search Algorithm}
\author{Shantanav Chakraborty}
\email{shantanav_with_u@iitj.ac.in}
\author{Satyabrata Adhikari}
\email{satya@iitj.ac.in}
\affiliation{Indian Institute of Technology Jodhpur, Jodhpur-342011, India}
\begin{abstract}
Entanglement lies at the heart of quantum mechanics and has no classical analogue. It is central to the speed up achieved by quantum algorithms over their classical counterparts. The Grover's search algorithm is one such algorithm which enables us to achieve a quadratic speed up over any known classical algorithm that searches for an element in an unstructured database. Here, we analyse and quantify the effects of entanglement in the generalized version of this algorithm for two qubits. By 'generalized', it is meant that the use of any arbitrary single qubit unitary gate is permitted to create superposed states. Our analysis has been firstly on a noise free environment and secondly in the presence of noise. In the absence of noise, we establish a relation between the concurrence and the amplitude of the final state thereby showing the explicit effects of entanglement on the same. Moreover, the effects of noisy channels, namely amplitude and phase damping channels are studied. We investigate the amount of quantum correlation in the states obtained after the  phase inversion stage of the algorithm followed by interaction of those states with the noisy environment. The quantum correlations are quantified by geometric discord. It has been revealed that the states generated after the effect of amplitude damping on the phase inverted states of the quantum search algorithm possess non-zero quantum correlation even when entanglement is absent. However, this is absent in the phase damping scenario. 
\end{abstract}
\maketitle
\section{Introduction}
Ever since its inception, quantum computations have opened up a new avenue for fast computation. Quantum computing involves the use of quantum mechanics to solve computational problems with the aim of achieving a significant speed up over classical means. At the heart of quantum computing lie two quantum algorithms which were instrumental in the development of this area. Peter Shor came up with an algorithm that could factor a number into two primes in polynomial time \cite{shor94}. It is still unknown whether this is achievable classically but the fact that the solution is hard has been acknowledged and hence cryptographic algorithms such as the RSA, has been built over this problem. Needless to say, with the advent of a quantum computer, our entire cryptosystem would be at risk. The other famous quantum algorithm is the Grover's search algorithm, developed by Lov Grover in 1996 \cite{Gro96}. This algorithm provides a quadratic speed up over the best known classical algorithm. The algorithm involves searching for one or more solutions from an unstructured database consisting of an exponential number of entries. While the best known classical algorithm would require $O(N)$ number of steps to achieve this task, a quantum computer equipped with the Grover's Algorithm would take $O(\sqrt{N})$ number of steps \cite{Gro96}. The Grover's algorithm has been shown to be a special case of the more general amplitude amplification algorithm \cite{Hoy02}which opens up an avenue for a large number of quantum algorithms not involving measurement.\\
In all such quantum algorithms, it has been debated that entanglement plays a crucial role in achieving a speed up over classical algorithms. In fact, it has been shown that entanglement is necessary to achieve an exponential speed up in Shor's Algorithm\cite{JL03}. In Grover's \cite{Gro96}, Deutsch-Jozsa \cite{DJ92} and Simon's algorithms \cite{Sim94},the presence of entanglement has been detected \cite{JL03, Bru10}. The number of such entangled states increase with the increase in the number of qubits \cite{Bru10}. The features of such entangled states are not known to the best of our knowledge. There may also exist states that may not be entangled but may possess some other quantum correlations and even such states are thought to be useful for quantum computation \cite{Datta08}.\\
In spite of all its progresses, one major setback in the implementation of a quantum computer has been the possibility of decoherence \cite{Pelli95}. On interaction with the environment, a quantum system may lose its coherence and this results in a loss of the quantum superposition of states thereby reducing to a classical state. Quantum circuits too are susceptible to decoherence and hence so are quantum algorithms.The effect of noise leads to erroneous results and hence must be accounted for \cite{Sho95}.\\ 

%It would be interesting however to study the effects of noise on the amount of entanglement in quantum algorithms as they lead to the creation of mixed %states. 

Let us now discuss the Grover's search algorithm. The Grover's algorithm involves solving the unordered search problem. The unordered search problem requires to find an index $i$ out of an unordered set of $N=2^n$ arbitrary strings $S=\{x_0, x_1, x_2,......x_n\}$ such that $x_i=1$ or in other words it searches for a solution. The Grover's Algorithm achieves this very task in $O(\sqrt{N})$ time whereas the best known classical algorithm requires $O(N)$ steps. At the end of the algorithm, we obtain the the solution with high probability. \\
Initially, the input states are prepared in $|0\rangle^{\otimes n}$ and the Hadamard gate is applied to the  first $n$ qubits as shown in Fig. 1 resulting in an equal superposition of all $N$ basis states $|\psi_1\rangle$. 
\begin{equation}
|\psi_1\rangle=\frac{1}{\sqrt{2^n}}\sum_{x=0}^{2^n-1}|x\rangle
\end{equation}
The target qubit is kept in the $|-\rangle$ state. This initiation phase is followed by the phase inversion stage which involves the application of an Oracle $O$ which consists of a function $f(x)$ such that \[
 f(x) =
  \begin{cases}
   0 & \text{if } x \text{ is a not a solution} \\
   1       & \text{if } x \text{ is a solution}
  \end{cases}
\]  
Thus $O$ flips the phase of only the solution without affecting the other states resulting in $|\psi_2\rangle$.
\begin{equation}
|\psi_2\rangle=\frac{1}{\sqrt{2^n}}\sum_{x=0}^{2^n-1}(-1)^{f(x)}|x\rangle
\end{equation}

The next stage is the application of the diffusion matrix $D=-I+[2/N]$ to the $n$ qubits. Here, $I$ is the $N\times N$ identity matrix while $[2/N]$ is the $N\times N$ matrix which each entry being $2/N$. 
\begin{equation}
D=2|0^n\rangle\langle0^n|-I_n
\end{equation}
The role of $D$ is to invert the states about the mean of the amplitude of the superposition of states. This stage, known as the inversion about mean amplifies the amplitude of the solution. The 'phase inversion' and the 'inversion about mean' stages together comprise of the Grover's iterate $G$ which is iterated $O(\sqrt{N})$ times.
\begin{figure}[h]
\centering
\begin{align*}
 \Qcircuit @C=1em @R=.7em {   
  &						&								&				 & & &\mbox{\textbf{Repeat $O(\sqrt{N})$ times}} & & \\
  & \lstick{|0\rangle} & /^n \qw & \gate{H^{\otimes n}} & \multigate{1}{U} & \gate{H^{\otimes n}} & \gate{2 |{0^n}\rangle \langle{0^n}| - I_n}         & \gate{H^{\otimes n}} & \qw  \\
 & \lstick{|1\rangle} & \qw     & \gate{H}     & \ghost{U}        & \qw                  & \qw     &\qw   &\qw \qw \gategroup{2}{5}{3}{8}{.7em}{--} \\                              
 }
 \end{align*}
\caption{Circuit for Grover's Algorithm}
\end{figure}
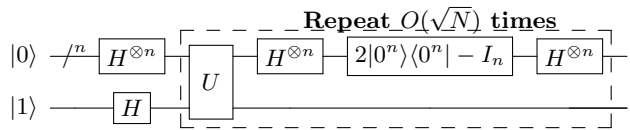 
\\
In this article, we have replaced the Hadamard Transformations with a general unitary gate $U$ and extensively studied the effects of noise, namely amplitude and phase damping, on quantum correlations in the generalized Grover's algorithm for two qubits. Now, there exists many stages where the system may decohere. In order to study the effects of entanglement, the most interesting would be the effect of noise just after entanglement is created. Thus in our case, we consider the effect of noise just after the application of Oracle $O$ on the superposition of basis states. We reveal the possibility of states having non zero quantum discord yet zero concurrence and the effect that such states would have on the algorithm. The effects of noise on the Grover's algorithm has been studied extensively \cite{Elli04}. The effect of noise in the performance of Grover's algorithm has been studied in \cite{Shap03, Norm99, Gaw11}. In section I, the effects of entanglement in this algorithm without noise, has been studied. Here, an explicit relation between concurrence and amplitude has been established. In Section II and Section III, we study the effects of amplitude and phase damping noisy channels on this algorithm. Finally, we conclude in section IV. 

\section{Entanglement in generalized quantum search algorithm for two qubits in a noise free environment} 
Let us assume that in the initiation phase of the Grover's algorithm a generalized Unitary gate $U$ has been used to create a superposition of states. The transformation of $U$ is given by:
\begin{equation}
U|0\rangle=\alpha|0\rangle +\beta|1\rangle \text{ and, }U|1\rangle=\beta|0\rangle - \alpha|1\rangle
\end{equation}
such that $\alpha^2+\beta^2=1$.
 Without any loss of generality we assume that $\alpha$ and $\beta$ are real. Thus, the initial state which is in $|00\rangle$, changes to:
\begin{equation}
U^{\otimes 2}|00\rangle =|\psi_1\rangle = \alpha^2|00\rangle+\alpha \beta
|01\rangle+\alpha \beta|10\rangle+\beta^2|11\rangle 
\end{equation}
Next comes the phase inversion stage wherein the Oracle $O$ acts on $|\psi_1\rangle$. Without any loss of generality we assume that $f(01)=1$. The results obtained for the other three cases are similar. Thus, after the phase inversion stage:
\begin{equation}
|\psi_2\rangle= \alpha^2|00\rangle-\alpha\beta|01\rangle+\alpha\beta|10\rangle+\beta^2|11\rangle 
\end{equation}
The concurrence, $C(\psi)$, of a pure state $|\psi\rangle$ is an entanglement monotone which allows us to quantify entanglement in such states\cite{Woo98}. Concurrence holds equally good for mixed states as well which we shall define in the subsequent sections. For pure states,  
\begin{equation}
C(\psi)=|\langle\psi|\tilde{\psi}\rangle| \text{, } |\tilde{\psi}\rangle= \sigma_y|\psi^*\rangle
\end{equation} 
where, $|\psi^*\rangle$ is the complex conjugate of the state $|\psi\rangle$ and $\sigma_y$ denote the Pauli spin matrix. In our case, $|\psi\rangle=|\psi^*\rangle$ as coefficients are real. In our case, 
\begin{equation}
C(|\psi_2\rangle)= c = 4\alpha^2\beta^2 
\end{equation}
Thus, $\alpha^2=\frac{1\pm\sqrt{1-c}}{2}$ and $\beta^2=\frac{1\mp\sqrt{1-c}}{2}$. State $|\psi_2\rangle$ can be re-expressed in terms of $c$ as:
\begin{equation}
|\psi_2\rangle=\frac{1}{2}[(1\pm \sqrt{1-c}|00\rangle - \sqrt{c}|01\rangle + \sqrt{c}|10\rangle + 1\mp \sqrt{1-c}|11\rangle] 
\end{equation} 
Clearly, $c=1$ corresponds to the Grover's algorithm with $U$ being the Hadamard Transform $H$, such that there exists an equal superposition of the states. This also indicates that the entanglement is maximal in such states. \\
For the generalized Grover algorithm, the diffusion matrix is expressed in terms of the unitary transform $U$ as $D=UA_0U$, where, $A_0$ is the conditional phase shift operator, inverting the phase of every computational basis state other than $|0\rangle$. Thus,
\begin{equation}
D=
\left[\begin{smallmatrix}
\alpha^4 - 2\alpha^2\beta^2-\beta^4 & 2\alpha^3\beta & 2\alpha\beta & 2\alpha^2\beta^2 \\
2\alpha^3\beta & -\alpha^4-\beta^4 & 2\alpha^2\beta^2 & 2\alpha\beta^3\\
2\alpha^3\beta & 2\alpha^2\beta^2 & -\alpha^4-\beta^4 & 2\alpha\beta^3\\
2\alpha^2\beta^2 & 2\alpha\beta^3 & 2\alpha\beta^3\ & -\alpha^4 - 2\alpha^2\beta^2+\beta^4  
\end{smallmatrix}\right]
\end{equation}
In terms of concurrence $c$, the diffusion matrix is obtained as:  
\begin{equation}
D=\frac{1}{8}
\left[\begin{smallmatrix}
\pm\sqrt{1-c} - c/2 & \frac{1}{2}\sqrt{c}(1\pm\sqrt{1-c}) & \frac{1}{2}\sqrt{c}(1\pm\sqrt{1-c}) & c/2 \\
\frac{1}{2}\sqrt{c}(1\pm\sqrt{1-c})& \frac{c-2}{2} & c/2 & \frac{1}{2}\sqrt{c}(1\pm\sqrt{1-c})\\
\frac{1}{2}\sqrt{c}(1\pm\sqrt{1-c}) & c/2 & \frac{c-2}{2} & \frac{1}{2}\sqrt{c}(1\mp\sqrt{1-c})\\
c/2 & \frac{1}{2}\sqrt{c}(1\mp\sqrt{1-c}) & \frac{1}{2}\sqrt{c}(1\pm\sqrt{1-c}) & \mp\sqrt{1-c} - c/2  
\end{smallmatrix}\right]
\end{equation}
And finally, the state after the application of $D$ is:
\begin{equation}
|\psi_3\rangle= \frac{1}{8}
\left[\begin{smallmatrix}
(4-4c)\pm(4-4c)\sqrt{1-c}\\
12\sqrt{c}-4c \sqrt{c}\\
4\sqrt{c}-4c \sqrt{c}\\
(4-4c)\mp(4-4c)\sqrt{1-c}\\
\end{smallmatrix}\right]
\end{equation}

In order to observe how the amplitude of the solution varies with the amount of entanglement, we plot a graph showing the same (the plot corresponds to the amplitude of $|00\rangle=(4-4c)+(4-4c)\sqrt{1-c}$, the other case is exactly same with the amplitudes of $|00\rangle$ and $|11\rangle$ interchanged) in Fig.2. 
\begin{figure}[h]
\includegraphics[scale=0.25]{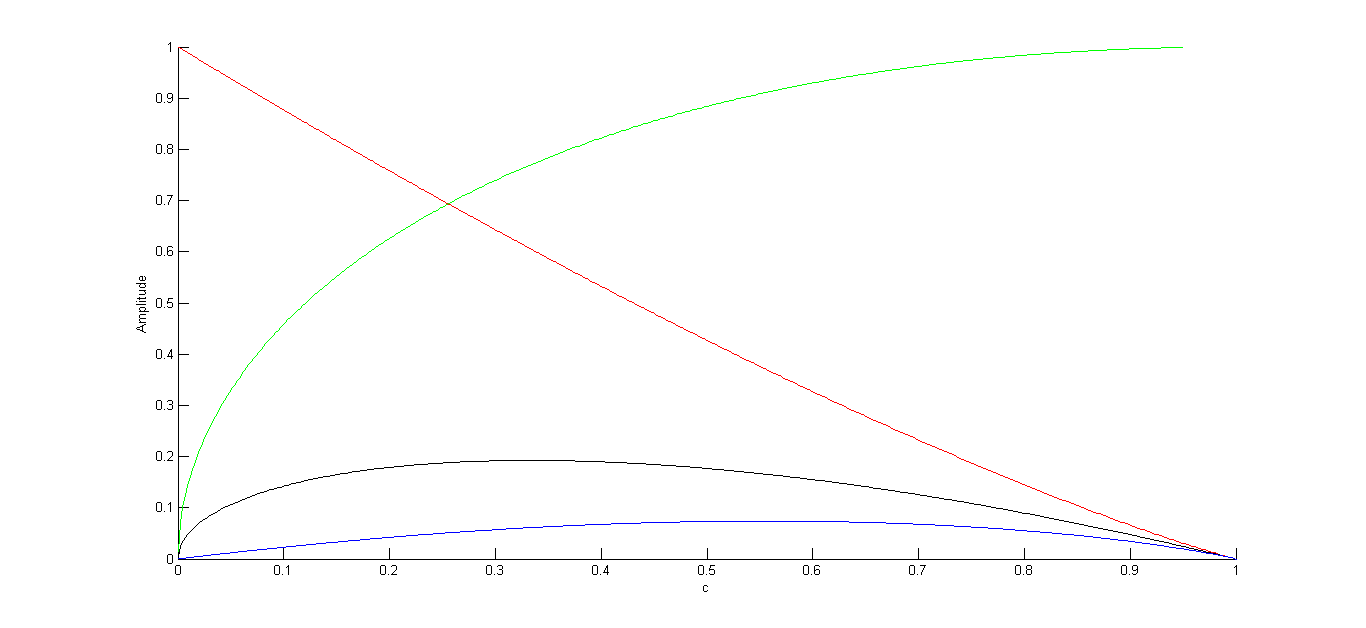}
\centering
\caption{Plot of amplitude against concurrence $c$ in the absence of noise. The green curve is the amplitude of $|01\rangle$, the red curve indicates $|00\rangle$, blue being $|10\rangle$ and black being $|11\rangle$}
\end{figure}
The graph shows that with the increase in entanglement, the amplitude of the solution increases and hence the probability of measuring the solution is maximum when $c=1$. However, it is interesting that a sufficient amount of concurrence is required to segregate the solution state from other states. In fact, after calculating we find that for $c\geq 0.256$, the probability of measuring the solution $|01\rangle$ is greater than that of measuring $|00\rangle$. Clearly, for proper functioning of the algorithm, a sufficient amount of entanglement must be present in the initial superposition of states. The best case scenario is when the Hadamard transform is applied to create an equal superposition of states. The probability of measuring the solution is maximum in that case.\\
Now the cases where $f(00)$, $f(10)$  or $f(11)$ are individually $1$, are similar and all such states become maximally entangled after the application of the oracle $O$. The cases where there exists two solutions are uninteresting as there does not exist any entanglement. For three solutions, the situation is similar to the case where exactly one solution exists differing only by a global phase. Thus for two qubits, after the application of the oracle the states are either maximally entangled or are product states. Let us now analyse the effects of the amplitude and phase damping channels on the amount of entanglement.
\section{Effect of Amplitude Damping on the Quantum search algorithm}
The amplitude damping channel on a qubit transforms the state $|1\rangle$ to the state $|0\rangle$ with a probability $p$ and keeps the state $|0\rangle$ unchanged \cite{NieChu00}. The Kraus Operators for the amplitude damping channel are:
\begin{equation}
E_0=\left[\begin{smallmatrix}
1 & 0\\
0 & \sqrt{1-p}
\end{smallmatrix}\right]
\text{ and }
E_1= \left[\begin{smallmatrix}
1 & \sqrt{p}\\
0 & 0
\end{smallmatrix}\right]
\end{equation}
Now, in the generalized search algorithm described in section II, the effect of noise due to amplitude damping has been taken into consideration just after the application of the Oracle as shown in Fig. 2. In our case, the second qubit of the state $|\psi_2\rangle$ gets affected. One may also study the effects on the first or both the qubits. However, we concentrate on the effects of noise in the second qubit only. In the density operator notation, $|\psi_2\rangle$ is expressed as:
\begin{equation}
\rho_{\psi_2}=\left[
\begin{smallmatrix}
\alpha^4 & -\alpha^3\beta & \alpha^3\beta & \alpha^2\beta^2\\
-\alpha^3\beta & \alpha^2\beta^2 & -\alpha^2\beta^2 & \alpha\beta^3\\
\alpha^3\beta & -\alpha^2\beta^2 & \alpha^2\beta^2 & \alpha\beta^3\\
\alpha^2\beta^2 & -\alpha\beta^3 & \alpha\beta^3 & \beta^4 
\end{smallmatrix}\right]
\end{equation}
Clearly, if amplitude damping affects only the second qubit, the Kraus operators become $I\otimes E_0$ and $I\otimes E_1$ respectively. In the forthcoming discussions, whenever the Kraus operators are mentioned, the action of identity matrix $I$ on the first qubit is implied.
The effect of amplitude damping on the second qubit would change the state as:
\begin{equation}
\tilde{\rho_{\psi_2}}= E_0\rho_{\psi_2}{E_0^{\dag}} + E_1\rho_{\psi_2}{E_1^{\dag}}
\end{equation}
and thus,
\begin{equation}
\tilde{\rho_{\psi_2}}=
\left[\begin{smallmatrix}
\alpha^4+p\alpha^2\beta^2 & -\sqrt{1-p}\alpha^3\beta & \alpha^3\beta-p\alpha\beta^3 & \sqrt{1-p}\alpha^2\beta^2\\
-\sqrt{1-p}\alpha^3\beta & (1-p)\alpha^2\beta^2 & -\sqrt{1-p}\alpha^2\beta^2 & -(1-p)\alpha\beta^3\\
\alpha^3\beta-p\alpha\beta^3 & -\sqrt{1-p}\alpha^3\beta & \alpha^2\beta^2+p\beta^4 & \sqrt{1-p}\alpha\beta^3\\
\sqrt{1-p}\alpha^2\beta^2 & -(1-p)\alpha\beta^3 & \sqrt{1-p}\alpha\beta^3 & (1-p)\beta^4   
\end{smallmatrix}\right]
\end{equation}
To find out the concurrence of $\tilde{\rho_{\psi_2}}$, we use the definition of concurrence for a mixed state $\rho$ given by $C(\rho)=max\{0,\lambda_0-\lambda_1-\lambda_2-\lambda_3\}$ where $\lambda_0, \lambda_1, \lambda_2, \lambda_3$ are the square roots of the eigenvalues of the matrix $\rho\rho^{\dag}$ arranged in decsending order and $\rho^\dag=\sigma_y\otimes\sigma_y\rho^*\sigma_y\otimes\sigma_y$ with $\rho^*$ being the complex conjugate of $\rho$ \cite{Woo98, Woo01}. For $\tilde{\rho_{\psi_2}}$, the concurrence comes out to be $C=max\{0,4\alpha^2\beta^2\sqrt{1-p}\}$. Thus, $C= 4\alpha^2\beta^2\sqrt{1-p}$. Clearly, when $\alpha=\beta=\frac{1}{\sqrt{2}}, C=\sqrt{1-p}$. The variation of concurrence with the amplitude $\alpha$ and the probability $p$ is shown in Fig. 3. Concurrence is maximum when $p=0$, indicating that the states are maximally entangled similar to the case in section I. 
\begin{figure}[h]
\includegraphics[scale=0.4]{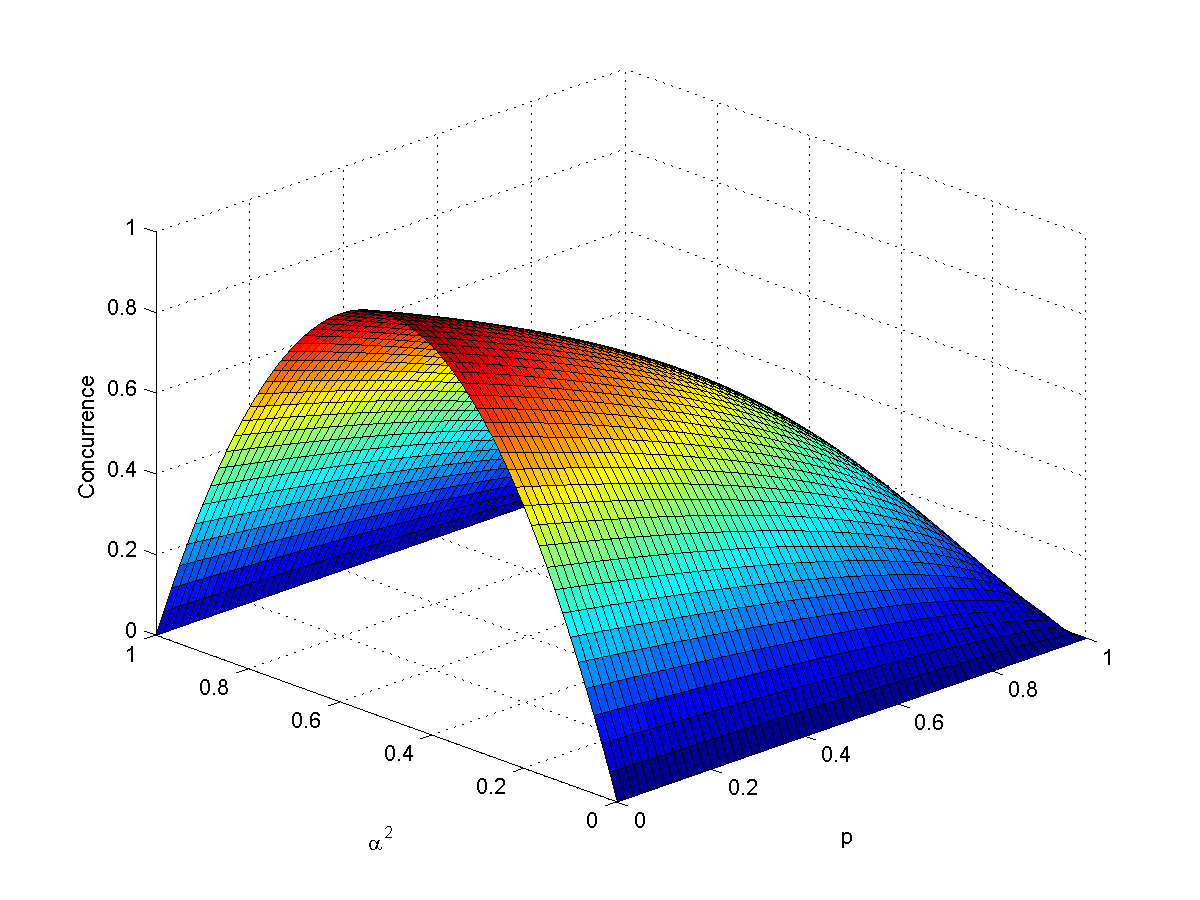}
\centering
\caption{Surface plot of Concurrence $(c)$ against probability of noise affecting the qubit $(p)$ and $(\alpha^2)$ for amplitude damping channel}
\end{figure}

On the other hand, when $p=1$, there exists no entanglement as $c=0$ and the situation is 'noisiest' ($c$ may also be zero when either or both of $\alpha$ and $\beta$ are zero). Thus, for $p=1$, we find that the entanglement which was maximal after the application of the Oracle is now lost. However, it is interesting to study whether there exists any other form of quantum correlation other than entanglement. When $p=1$, concurrence drops to zero indicating that there exists no entanglement. However, as we shall see later, that although the states become separable, there exists some form of quantum correlation as is indicated by non-zero value of geometric discord \cite{Bruk10, Giro11}. Recently a relationship has been established between geometric discord and teleportation fidelity \cite{Adhi13}.
\\
Now a general two qubit density matrix is given by:
\begin{equation}
\rho=\frac{1}{4}[I\otimes I + \vec{r}.\vec{\sigma}\otimes I + I\otimes \vec{s}.\vec{\sigma}+\sum_{i,j}t_{ij}\sigma_i\otimes\sigma_j] 
\end{equation}
Here, $I$ is the $2\times2$ identity matrix, $x_i=Tr(\rho(\sigma_i\otimes I_2)$  and $y_i= Tr(\rho(I_2\otimes \sigma_i)$ are the Bloch Sphere components and ${t_{ij}}\equiv Tr(\rho(\sigma_i\otimes \sigma_j))$ represents the element of the correlation matrix $T$ with $\sigma_i$ denoting the Pauli Matrices for $i \in \{1,2,3\}$. Next, we study the quantum correlations in the state $\tilde{\rho_{\psi_2}}$.
For this we express, $\tilde{\rho_{\psi_2}}$ in the above form and compare the coefficients with the above equation, we get the following values for the coefficients:
$r_x=2\alpha^3\beta-2\alpha\beta^3; r_y=0; r_z=\alpha^4-\beta^4; s_x=2\sqrt{1-p}(\alpha\beta^3-\alpha^3\beta); s_y=0; s_z=\alpha^4+2p\beta^2-\beta^4; t_{xx}=t_{xy}=t_{yz}=t_{yx}=t_{yz}=t_{zy}=0$; $t_{xz}=2\alpha\beta(1-2p\beta^2); t_{yy}=4\alpha^2\beta^2\sqrt{1-p}$; $t_{zx}=-4\alpha^3\beta\sqrt{1-p}$; $t_{zz}=(\alpha^2-\beta^2){\alpha^2+(2p-1)\beta^2}$.
\\
The geometric discord quantifies the amount of quantum correlation for a two qubit system, and has been shown to be $D_G(\rho)=\frac{1}{2}(\|\vec{x}\|^2 + \|T\|^2 - k_{max})$ where, $\|.\|_2$ denotes the Hilbert Schmidt norm defined as $\|A\|_2=\sqrt{Tr(A^{\dag} A)}$ and $k_{max}$ denotes the maximum eigenvalue of the matrix $(\vec{x}\vec{x}^{\dag}+TT^{\dag})$ \cite{Bruk10}. Thus for the state $\rho_{\psi_2}$, the value of discord, $D_G(\tilde{\rho_{\psi_2}})= \frac{1}{2}[\frac{1}{2}\alpha^8 + \frac{1}{2}\beta^8 + \frac{1}{2}\alpha^4+\frac{1}{2}\beta^4-2p\beta^8+\alpha^2\beta^2+15\alpha^4\beta^4+8\alpha^6\beta^2+2p^2\beta^8-8p\alpha^2\beta^4+6p\alpha^2\beta^6-22p\alpha^4\beta^4-6p\alpha^6\beta^2+ 4p^2\alpha^2\beta^6+2p^2\alpha^4\beta^4-(\alpha^{16}-24p\alpha^{14}\beta^2+16\alpha^{14}\beta^2+152p^2\alpha^{12}\beta^4-216p\alpha^{12}\beta^4+92\alpha^{12}\beta^4+2\alpha^{12}
-96p^3\alpha^{10}\beta^6+368p^2\alpha^{10}\beta^6-504p\alpha^{10}\beta^6+208\alpha^{10}\beta^6-96p\alpha^{10}\beta^4-24p\alpha^{10}\beta^2+68\alpha^{10}\beta^2+16p^4\alpha^8\beta^8
+736p^3\alpha^8\beta^8-664p^2\alpha^8\beta^8+40p\alpha^8\beta^8+70\alpha^8\beta^8-768p^2\alpha^8\beta^6+768p\alpha^8\beta^6+8p^2\alpha^8\beta^4+248p\alpha^8\beta^4
-322\alpha^8\beta^4+\alpha^8+64p^4\alpha^6\beta^{10}-192p^3\alpha^6\beta^{10}+480p^2\alpha^6\beta^{10}+440p\alpha^6\beta^{10}-208\alpha^6\beta^{10}-128p^3\alpha^6\beta^8
+768p^2\alpha^6\beta^8-320p\alpha^6\beta^8-32p^2\alpha^6\beta^6-240p\alpha^6\beta^6+32p\alpha^6\beta^4-12\alpha^6\beta^2+96p^4\alpha^4\beta^{12}+64p^3\alpha^4\beta^{12}
+488p^2\alpha^4\beta^{12}-328p\alpha^4\beta^{12}+92\alpha^4\beta^{12}-256p^3\alpha^4\beta^{10}-768p^2\alpha^4\beta^{10}+256p\alpha^4\beta^{10}+432p^2\alpha^4\beta^8
+208p\alpha^4\beta^8-130\alpha^4\beta^8-192p\alpha^4\beta^6+38\alpha^4\beta^4+64p^4\alpha^2\beta^{14}+32p^3\alpha^2\beta^{14}-144p^2\alpha^2\beta^{14}+88p\alpha^2\beta^{14}
-16\alpha^2\beta^{14}-128p^3\alpha^2\beta^{12}+256p^2\alpha^2\beta^{12}-96p\alpha^2\beta^{12}-32p^2\alpha^2\beta^{10}-56p\alpha^2\beta^{10}+36\alpha^2\beta^{10}+32p\alpha^2\beta^8
-12\alpha^2\beta^6+16p^4\beta^{16}-32p^3\beta^{16}+24p^2\beta^{16}-8p\beta^{16}+\beta^{16}+8p^2\beta^{12}-8p\beta^{12}+2\beta^{12}+\beta^8)]$
Fig. 4 depicts the variation in discord. 
\begin{figure}[h]
\includegraphics[scale=0.4]{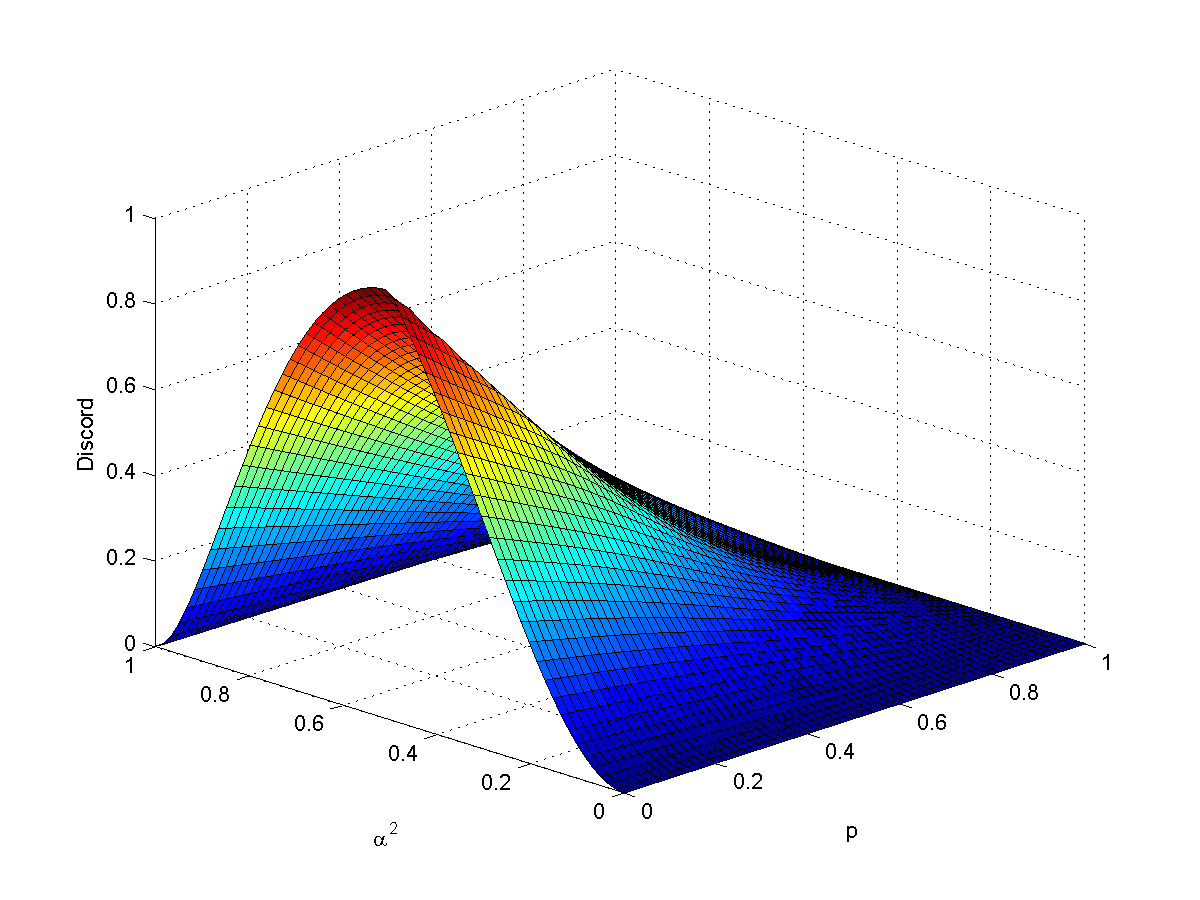}
\centering
\caption{Surface plot of quantum discord against probability of noise affecting the qubit $(p)$ and $(\alpha^2)$ for amplitude damping}
\end{figure}
Earlier, we had observed that the value of concurrence goes to $0$ when $p=1$ and had argued that there may exist some non-classical correlation other than entanglement. This is confirmed in Fig. 5 which depicts the variation of discord against $\alpha^2$ when $p=1$. Discord is maximum when $\alpha=\frac{1}{\sqrt{2}}$, i.e. when $U=H$.
\begin{figure}[h]
\centering
\includegraphics[scale=0.25]{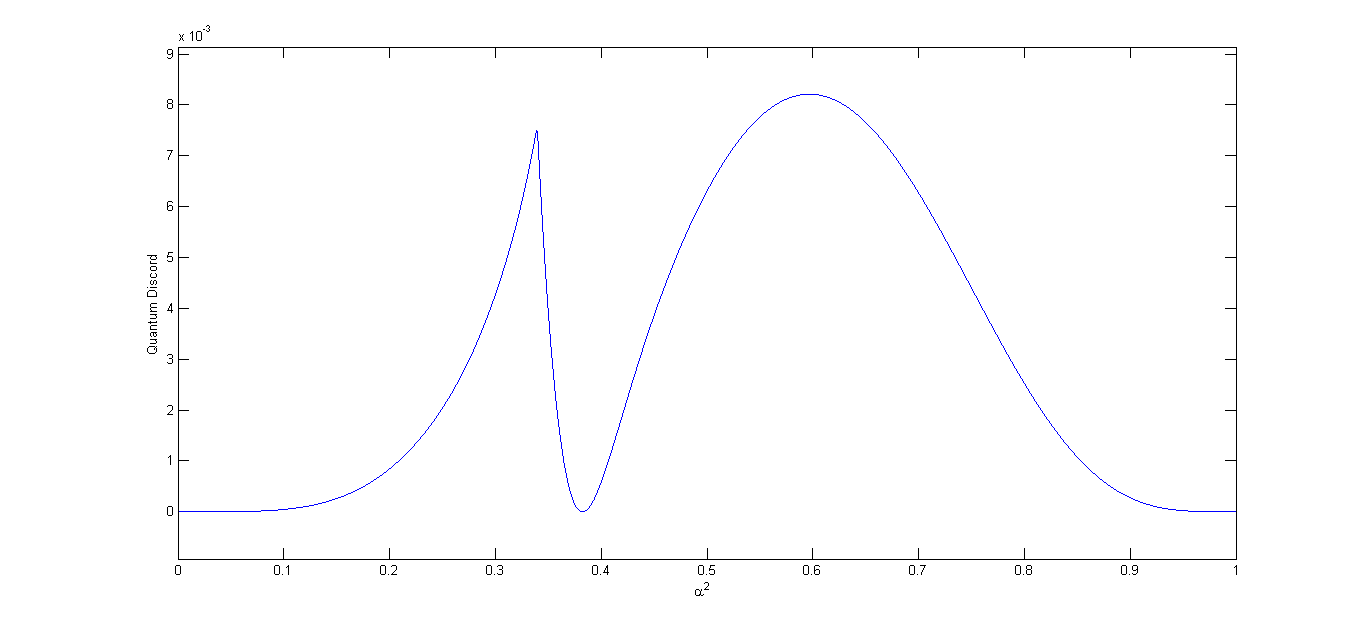}
\caption{Discord in the absence of entanglement $(p=1)$}
\end{figure}
\\
After the application of the diffusion matrix $D=UA_0U$ to $\tilde{\rho_{\psi_2}}$ , we measure the solution in the standard basis. The solution would of course have the highest amplitude and hence we shall measure it with the highest probability. We applied $D$ to the general two qubit density matrix $\rho$ (as shown in Equation 17) and calculate the same for amplitude and phase damping channels by substituting the necessary values. Thus for a general two qubit density matrix $\rho$, we obtain:

$A_1 = Tr[D\rho D|00\rangle\langle00|]= \frac{1}{4}\{t_{yy}(4\alpha^6\beta^2+8\alpha^4\beta^4+4\alpha^2\beta^6)+t_{xx}(-8\alpha^4\beta^4+12\alpha^6\beta^2-4\alpha^2\beta^6)+t_{zz}(4\alpha^2\beta^6-12\alpha^6\beta^2+6\alpha^4\beta^4+\alpha^8+\beta^8)+t_{xz}(4\alpha^7\beta-16\alpha^5\beta^3-4\alpha^3\beta^5)+ t_{zx}(4\alpha^7\beta-16\alpha^5\beta^3-4\alpha^3\beta^5)+(6\alpha^4\beta^4+4\alpha^6\beta^2+4\alpha^2\beta^6+\alpha^8+\beta^8)\}$,

$A_2 = Tr[D\rho D|01\rangle\langle01|]= \frac{1}{4}\{t_{yy}(-4\alpha^6\beta^2-8\alpha^4\beta^4-4\alpha^2\beta^6)+t_{xx}(-8\alpha^4\beta^4-4\alpha^6\beta^2-4\alpha^2\beta^6)+t_{zz}(4\alpha^2\beta^6+4\alpha^6\beta^2-6\alpha^4\beta^4-\alpha^8-\beta^8)+t_{xz}(4\alpha\beta^7+12\alpha^5\beta^3) +t_{zx}(-4\alpha^7\beta-12\alpha^3\beta^5)+(6\alpha^4\beta^4+4\alpha^6\beta^2+4\alpha^2\beta^6+\alpha^8+\beta^8)\}$,
 
$A_3 = Tr[D\rho D|10\rangle\langle10|]= \frac{1}{4}\{t_{yy}(-4\alpha^6\beta^2-8\alpha^4\beta^4-4\alpha^2\beta^6)+t_{xx}(8\alpha^4\beta^4-4\alpha^6\beta^2-4\alpha^2\beta^6)+t_{zz}(4\alpha^2\beta^6+4\alpha^6\beta^2-6\alpha^4\beta^4-\alpha^8-\beta^8)+t_{xz}(-4\alpha^7\beta-12\alpha^3\beta^5)+t_{zx}(4\alpha\beta^7+12\alpha^5\beta^3)+ (4\alpha^6\beta^2+4\alpha^2\beta^6+\alpha^8+\beta^8)\}$ and 
 
$A_4 = Tr[D\rho D|11\rangle\langle11|]= \frac{1}{4}\{t_{yy}(4\alpha^6\beta^2+8\alpha^4\beta^4+4\alpha^2\beta^6)+t_{xx}(-8\alpha^4\beta^4-4\alpha^6\beta^2+12\alpha^2\beta^6)+t_{zz}(-12\alpha^2\beta^6+4\alpha^6\beta^2+6\alpha^4\beta^4+\alpha^8+\beta^8)+t_{xz}(-4\alpha^7\beta+16\alpha^5\beta^3+4\alpha^3\beta^5)+ t_{zx}(-4\alpha^7\beta+16\alpha^5\beta^3+4\alpha^3\beta^5)+(6\alpha^4\beta^4+4\alpha^6\beta^2+4\alpha^2\beta^6+\alpha^8+\beta^8)\}$        

Now for the already calculated values of $t_{ij},r_i,s_j$ for the amplitude damping channel, we find the amplitude of the final state. The effect of amplitude damping on the final state has been shown in Fig.6. 
\begin{figure}[h]
\centering
\subfigure[Amplitude of $|00\rangle$]{\includegraphics[scale=0.2]{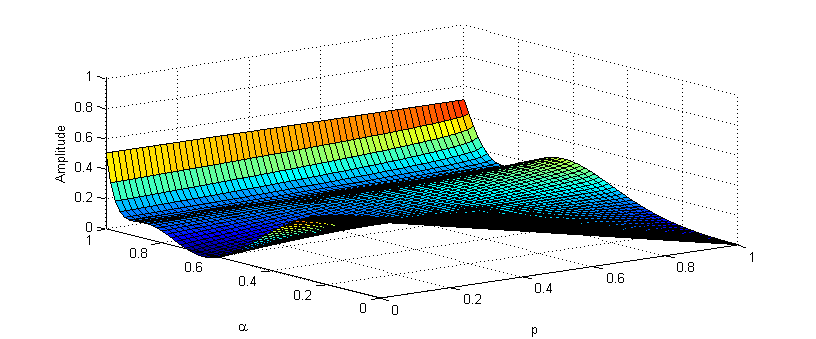}
}
\subfigure[Amplitude of $|01\rangle$]{\includegraphics[scale=0.2]{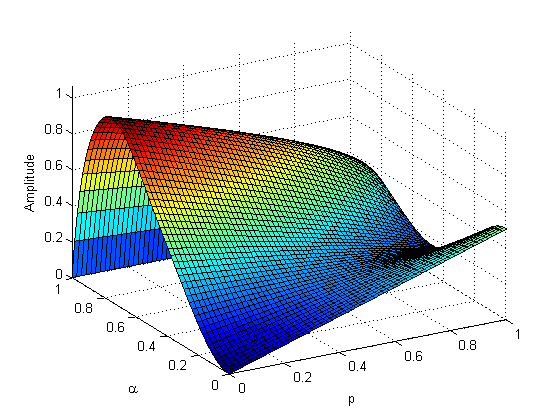}
}
\subfigure[Amplitude of $|10\rangle$]{\includegraphics[scale=0.15]{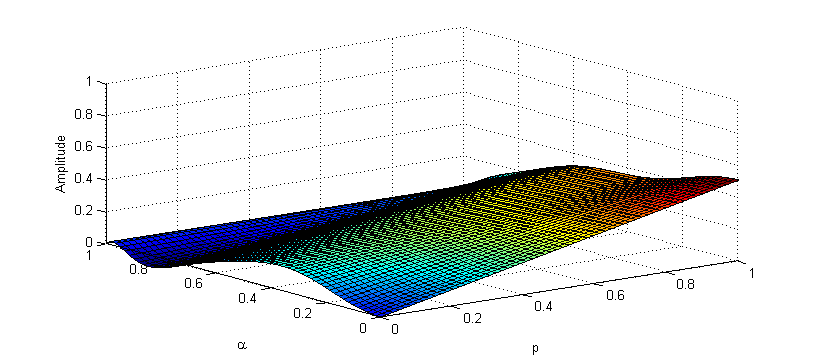}
}
\subfigure[Amplitude of $|11\rangle$]{\includegraphics[scale=0.15]{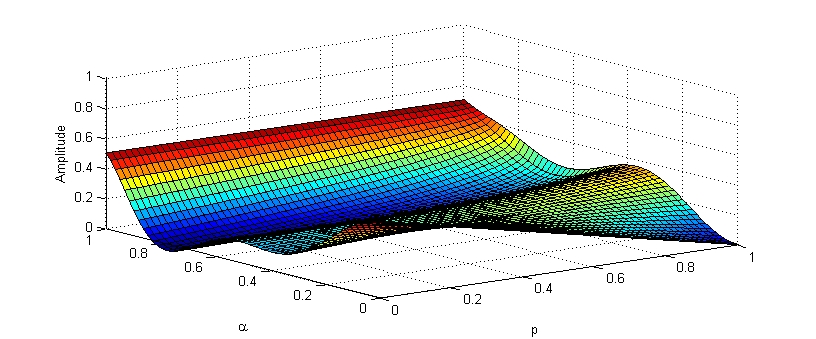}
}
\caption{Variation of amplitude of the basis states of the final state with $\alpha$ and $p$ for amplitude damping}
\end{figure}
The four plots correspond to the amplitude in the four basis states, namely $|00\rangle,|01\rangle,|10\rangle$ and $|11\rangle$. The amplitude of $|01\rangle$ remains maximum as it was the solution. However, the probability of measuring the same reduces as $p$ increases. When $U$ is the Hadamard gate, $\alpha=\frac{1}{\sqrt{2}}$ and this corresponds to the Grover's Algorithm. In that case, the variation of the amplitudes of the four computational basis states with $p$ is shown in Fig. 7.
\begin{figure}[h]
\includegraphics[scale=0.25]{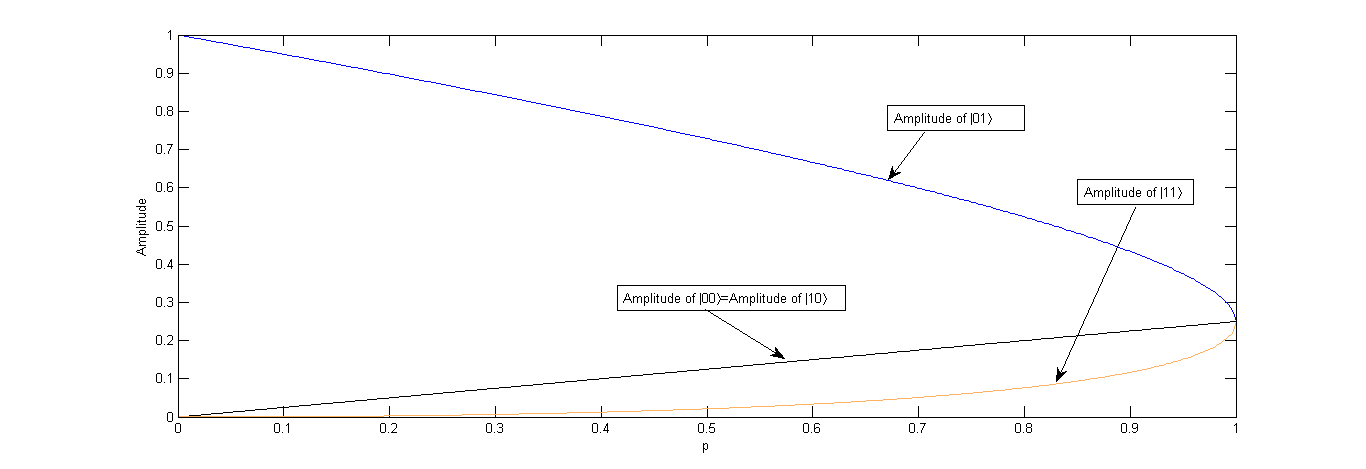}
\centering
\caption{Plot of amplitude against $p$ when $\alpha=\frac{1}{\sqrt{2}}$ for amplitude damping}
\end{figure}
The plot reveals that the probability of detecting the solution decreases gradually with the increase in $p$ and when $p=1$, the solution cannot be distinguished.

\section{Effect of Phase Damping on the quantum search algorithm}
The Kraus operators for the phase damping channel \cite{NieChu00} are:
$E_0=\left[\begin{smallmatrix}
1 & 0\\
0 & \sqrt{1-p}
\end{smallmatrix}\right]$
and, $E_1=\left[\begin{smallmatrix}
0 & 0\\
0 & \sqrt{p}
\end{smallmatrix}\right]$
Again, we assume that the effect of phase damping is on the second qubit and the Kraus operators are modified accordingly (by using $I\otimes E_0$ and $I\otimes E_1$). The resulting state $\tilde{\rho_{\psi_2}}= E_0\rho_{\psi_2}{E_0^{\dag}} + E_1\rho_{\psi_2}{E_1^{\dag}}$ which in matrix form is:
\begin{equation}
\tilde{\rho_{\psi_2}}=
\left[\begin{smallmatrix}
\alpha^4 & -\sqrt{1-p}\alpha^3\beta & \alpha^3\beta & \sqrt{1-p}\alpha^2\beta^2\\
-\sqrt{1-p}\alpha^3\beta & \alpha^2\beta^2 & -\sqrt{1-p}\alpha^2\beta^2 & -\alpha\beta^3\\
\alpha^3\beta -\sqrt{1-p}\alpha^2\beta^2 & \alpha^2\beta^2 & \sqrt{1-p}\alpha\beta^3\\
\sqrt{1-p}\alpha^2\beta^2 & -\alpha\beta^3 & \sqrt{1-p}\alpha^3\beta & \beta^4
\end{smallmatrix}\right]
\end{equation} 
The concurrence for $\tilde{\rho_{\psi_2}}$ is calculated. The concurrence $c=2\alpha^2\beta^2(\sqrt{2-p+2\sqrt{1-p}}-\sqrt{2-p-2\sqrt{1-p}})$. The effect of phase damping on the concurrence is depicted in Fig. 8. From the plot, we observe that $c=1$ when $\alpha=\frac{1}{\sqrt{2}}$ and $p=0$ (as is also indicated by the equation of concurrence). Concurrence goes to zero when $p=1$, i.e. the states would lose entanglement.
\begin{figure}[h]
\includegraphics[scale=0.4]{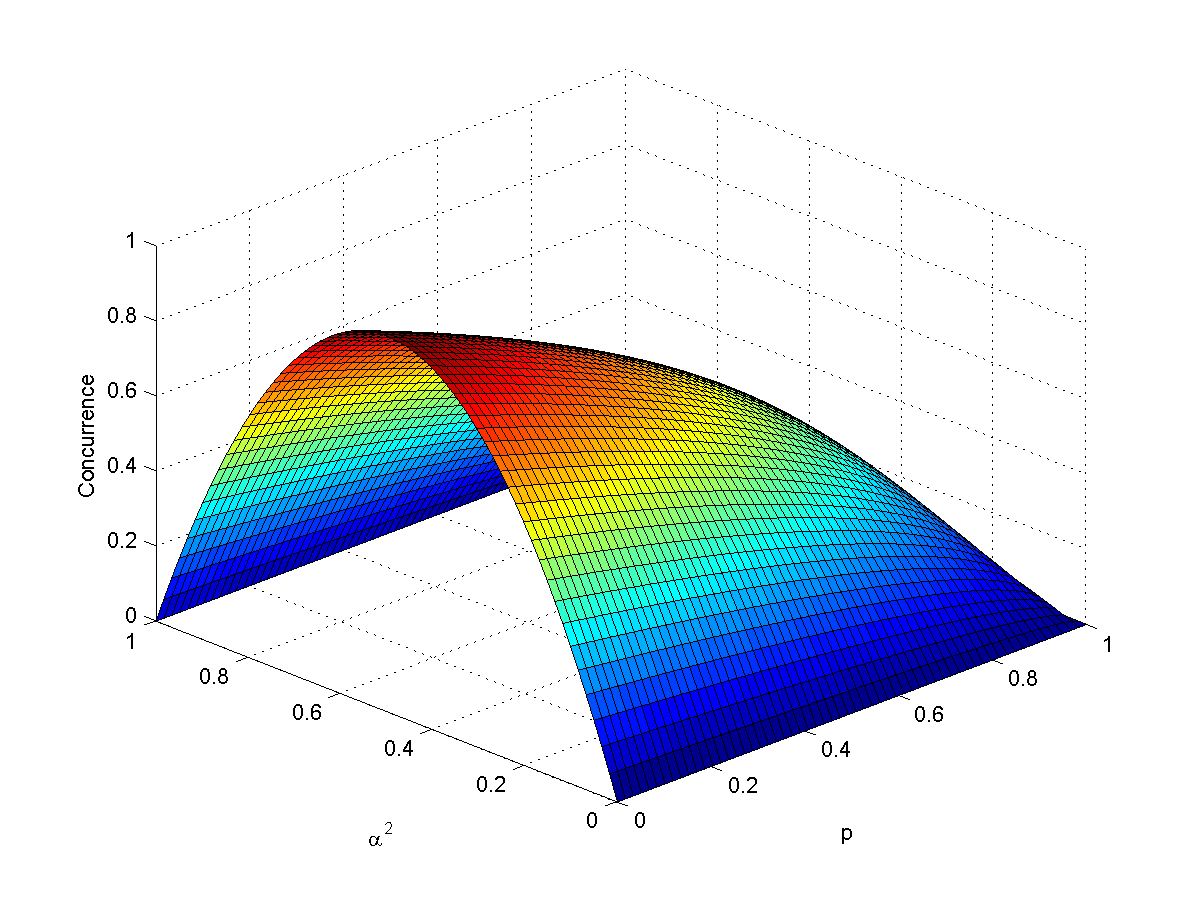}
\centering
\caption{Surface plot of Concurrence $(c)$ against probability of noise affecting the qubit $(p)$ and $(\alpha^2)$ for phase damping}
\end{figure}
\\
As in the case of the amplitude damping channel, we express the noise affected state in the form of a general a two qubit density matrix and compare the coefficients. After comparing the coefficients, the correlation matrix $T=\left[\begin{smallmatrix} 0 & 0 & 2\alpha^3\beta+2\alpha\beta^3\\
0 & -4\alpha^2\beta^2\sqrt{1-p} & 0\\ -2\sqrt{1-p}(\alpha^3\beta+\alpha\beta^3) & 0 & 0 
\end{smallmatrix}\right]$, the local Bloch sphere coefficient vector $X=\left[\begin{smallmatrix} 2\alpha^3\beta-2\alpha\beta^3\\0\\0 
\end{smallmatrix}\right]$. From the above two matrices, we obtain quantum discord of $\tilde{\rho_{\psi_2}}$ as 
\begin{equation}
D(\rho)=\frac{1}{2}4\alpha^2\beta^2(1-p)(\alpha^4+6\alpha^4\beta^4+\beta^4)
\end{equation}
Fig. 9 depicts the variation of discord with the amplitude $\alpha$. Thus, from the plot, we find that discord goes to zero when $p=1$, when concurrence is also zero. This rules out the possibility of any non-classical correlation other than entanglement as was the case in amplitude damping. Also, this shows that phase damping is more damaging as compared to amplitude damping, as in this case, noise reduces a highly entangled state to a mere classical state.
\begin{figure}[h]
\includegraphics[scale=0.4]{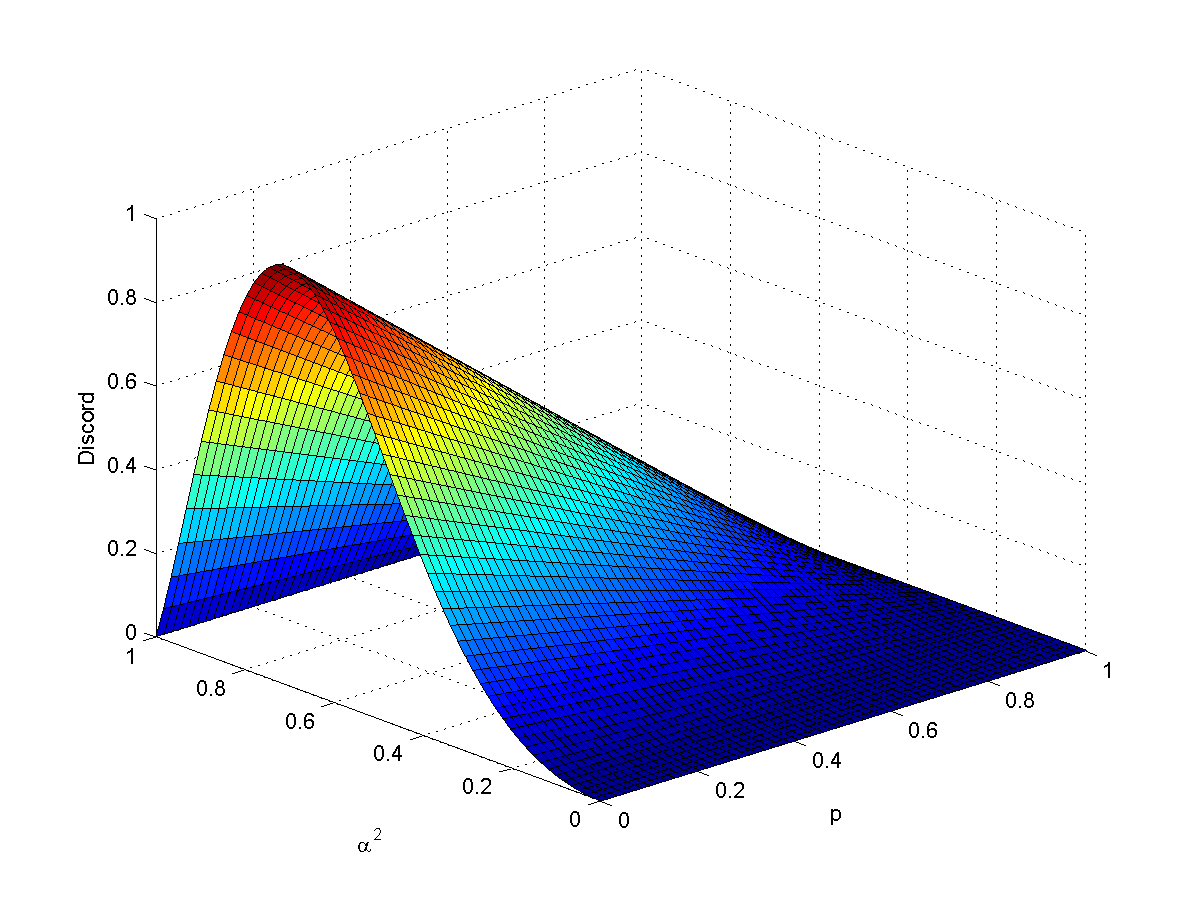}
\centering
\caption{Surface plot of quantum discord against probability of noise affecting the qubit $(p)$ and $(\alpha^2)$ for phase damping}
\end{figure}
The quantum discord is maximum, when $\alpha=\frac{1}{\sqrt{2}}$ and $p=0$ (as expected). The trend of the decrease in quantum discord with the increase in noise is shown in Fig. 10, showing clearly, that discord goes to zero as $p=1$.
\begin{figure}[h]
\includegraphics[scale=0.4]{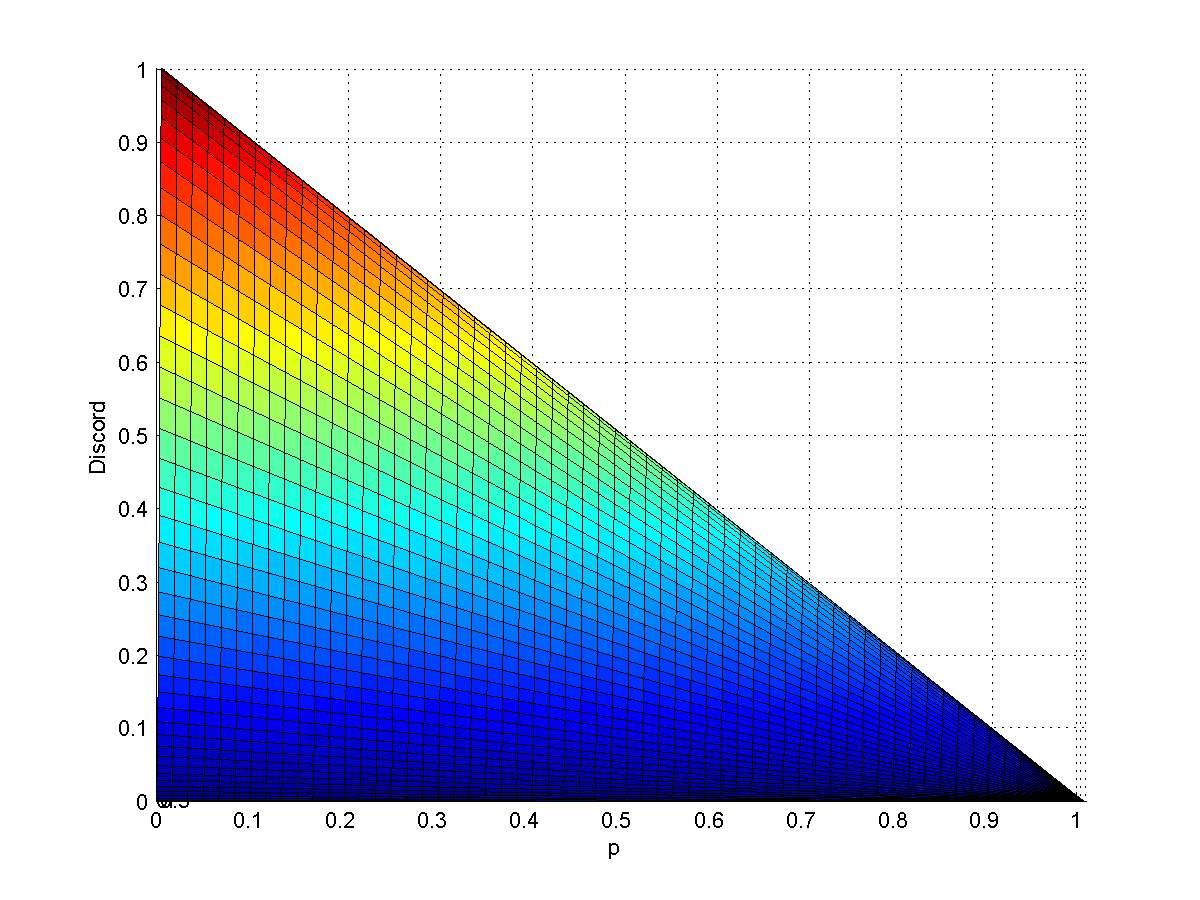}
\centering
\caption{Variation of discord with $p$ for phase damping}
\end{figure}  
Next, we analyse how the amplitude of the solution of the final states vary with the effect of phase damping. For this, the diffusion matrix $D$ acts on the state $\tilde{\rho_{\psi_2}}$ resulting in $\tilde{\rho_{\psi_3}}$. Thus, $\tilde{\rho_{\psi_3}}= D\tilde{\rho_{\psi_2}}D$  The amplitude of the state corresponding to the four basis states are obtained.
The amplitude of $|00\rangle=\frac{1}{4}[\alpha^8+\beta^8+6\alpha^4\beta^4+4\alpha^2\beta^6+4\alpha^6\beta^2]-\frac{1}{2}[\alpha\beta\{-4\alpha^7\beta+16\alpha^5\beta^3+4\alpha^3\beta^5\}]+\frac{1}{2}[\alpha\beta\sqrt{1-p}\{-4\alpha^7\beta+16\alpha^5\beta^3+4\alpha^3\beta^5\}]-\alpha^2\beta^2\sqrt{1-p}\{4\alpha^6\beta^2+8\alpha^4\beta^4+4\alpha^2\beta^6\}$, 
\\
\\
the amplitude of $|01\rangle=\frac{1}{4}[\alpha^8+\beta^8+6\alpha^4\beta^4+4\alpha^2\beta^6+4\alpha^6\beta^2]+\frac{1}{2}[\alpha\beta\{4\alpha\beta^7+12\alpha^5\beta^3\}]+\frac{1}{2}[\alpha\beta\sqrt{1-p}\{4\alpha^7\beta+12\alpha^3\beta^5\}]+\alpha^2\beta^2\sqrt{1-p}\{4\alpha^6\beta^2+8\alpha^4\beta^4+4\alpha^2\beta^6\}$, 
\\
\\
the amplitude of $|10\rangle=\frac{1}{4}[\alpha^8+\beta^8+6\alpha^4\beta^4+4\alpha^2\beta^6+4\alpha^6\beta^2]-\frac{1}{2}[\alpha\beta\{4\alpha^7\beta+12\alpha^3\beta^5\}]-\frac{1}{2}[\alpha\beta\sqrt{1-p}\{4\alpha\beta^7+12\alpha^5\beta^3\}]+\alpha^2\beta^2\sqrt{1-p}\{4\alpha^6\beta^2+8\alpha^4\beta^4+4\alpha^2\beta^6\}$ and, 
\\
\\
the amplitude of $|11\rangle=\frac{1}{4}[\alpha^8+\beta^8+6\alpha^4\beta^4+4\alpha^2\beta^6+4\alpha^6\beta^2]+\frac{1}{2}[\alpha\beta\{-4\alpha\beta^7+16\alpha^3\beta^5+4\alpha^5\beta^3\}]+\frac{1}{2}[\alpha\beta\sqrt{1-p}\{-4\alpha\beta^7+16\alpha^3\beta^5+4\alpha^5\beta^3\}]-\alpha^2\beta^2\sqrt{1-p}\{4\alpha^6\beta^2+8\alpha^4\beta^4+4\alpha^2\beta^6\}$.
The plots of the variation of the respective amplitudes are shown in Fig. 11. 
\begin{figure}[h]
\centering
\subfigure[Amplitude of $|00\rangle$]{\includegraphics[scale=0.2]{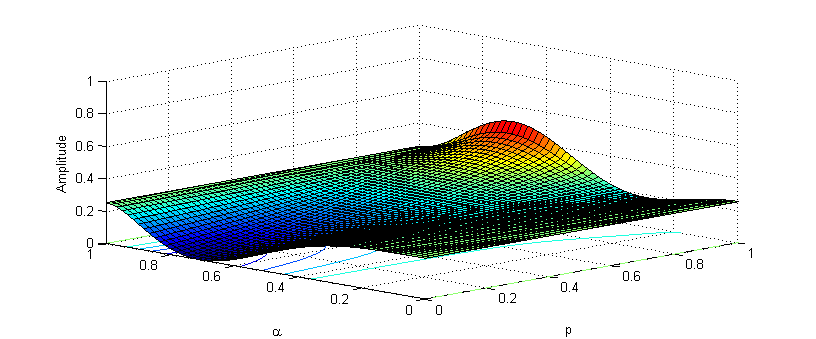}
}
\subfigure[Amplitude of $|01\rangle$]{\includegraphics[scale=0.2]{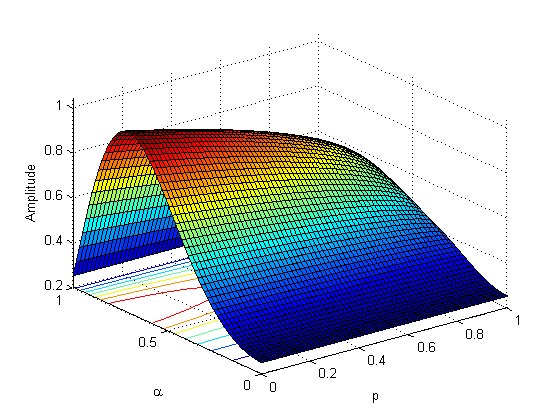}
}
\subfigure[Amplitude of $|10\rangle$]{\includegraphics[scale=0.15]{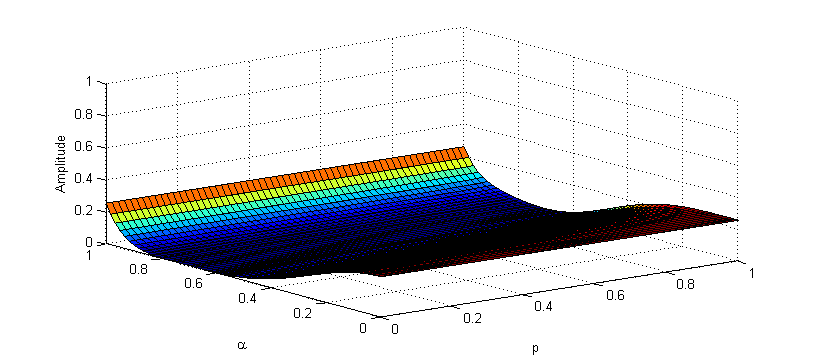}
}
\subfigure[Amplitude of $|11\rangle$]{\includegraphics[scale=0.15]{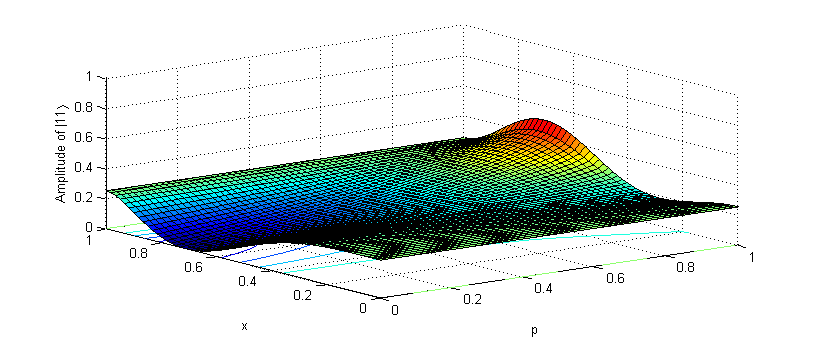}
}
\caption{Variation of amplitude of the basis states of the final state with $\alpha^2$ and $p$}
\end{figure}

The amplitude of $|11\rangle$ is the maximum (as $f(01)$ is considered to be $1$). For $\alpha=\frac{1}{\sqrt{2}}$, the variation of amplitude of $\tilde{\rho_{\psi_3}}$ has been plotted with change in $p$. 
\begin{figure}[h]
\includegraphics[scale=0.2]{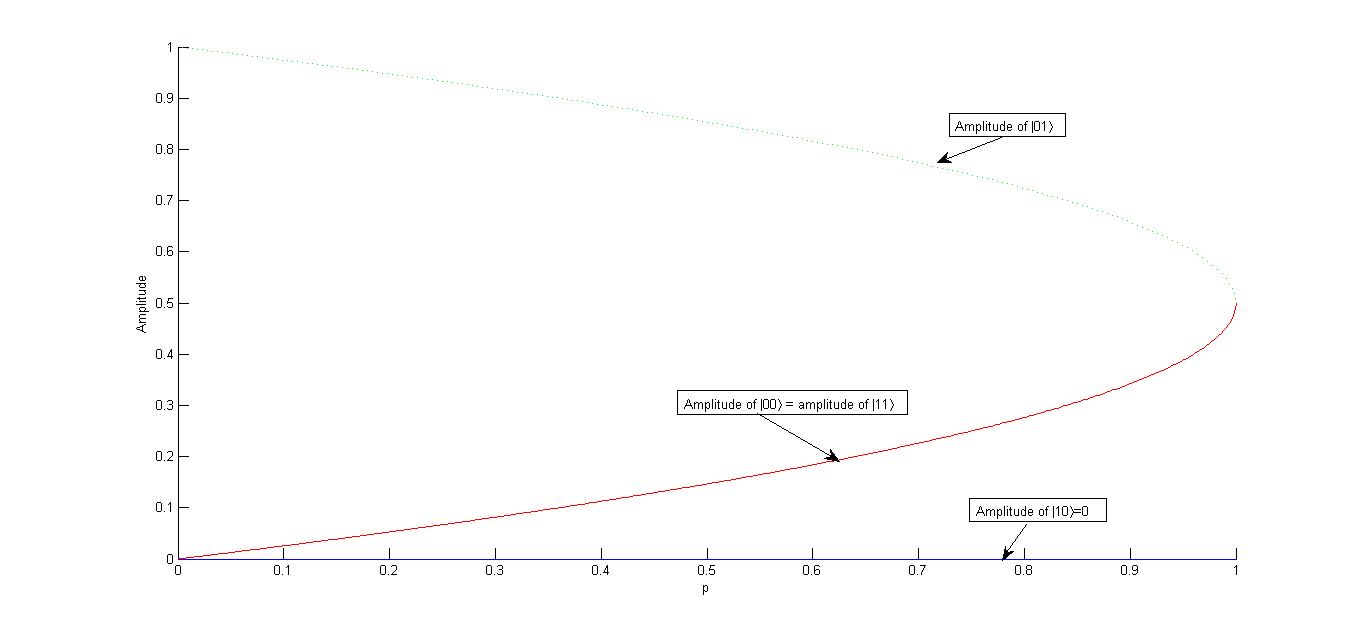}
\centering
\caption{Plot of amplitude against $p$ when $\alpha=\frac{1}{\sqrt{2}}$ for phase damping}
\end{figure}

We observe that for $p=1$, the amplitude of the solution $|01\rangle, |11\rangle$ and $|00\rangle$ coincide thereby making it impossible to detect the solution. Of course, with the increase in $p$, the probability of measuring the solution decreases and that of measuring the other states increase as shown in Fig. 12.

\section{Conclusion}

Our analysis of non-classical correlations of the generalized quantum search algorithm for two qubits can be divided into two parts. In the first part, we have shown that in the absence of noise, a certain amount of entanglement is required in order to amplify the amplitude of the solution sufficiently so as to have a high probability of measuring it. A relationship has been established between the amplitude of the final state and concurrence and it has been shown that the concurrence is maximum for the case where the Hadamard gate is used to create an initial superposition of states. In the second part, we have analysed the effects of phase damping and amplitude damping.Our study reveals how the amount of entanglement decreases with noise. We have also shown that it is possible for the mixed states obtained after the interaction with the amplitude damping channel, to be in some form of non-classical correlation other than entanglement as the geometric discord is non zero even in the absence of entanglement. Such a possibility, however is ruled out for the phase damping scenario. We have also shown how the final state of the algorithm are affected by noise. The nature of oracle used in the quantum search algorithm is similar to a variety of quantum algorithms including the Deutsch-Jozsa, Shor and Simon's algorithms and hence it would be interesting to carry out a similar analysis on the other algorithms too. Our immediate endeavours include analysing the entanglement in quantum search algorithms for multi-qubit systems \cite{Shan13}.  


\begin{thebibliography}{9}

\bibitem{shor94}
P.W.Shor, SIAM J. of Computing. \textbf{26}, 1484 (1997).
\bibitem{Gro96}
L.K. Grover, Phys. Rev. Lett. \textbf{79}, 325 (1997). 
\bibitem{Hoy02}
G. Brassard, P. Høyer, M. Mosca, Quant. Comp. and Quant. Info. Sc.,AMS Contemporary Mathematics Series. \textbf{305}, 53 (2002)
\bibitem{DJ92}
D. Deutsch and R. Jozsa, Proc. R. Soc. Lond. A \textbf{439}, 553 (1992).  
\bibitem{Sim94}
D. Simon, Proc. 35th IEEE Symposium on the Foundations of Computer Science, 116 (1994).
\bibitem{JL03}
R. Jozsa and N. Linden, Proc. R. Soc. Lond. A \textbf{459}, 2011 (2003).
\bibitem{Bru10}
D. Bruß, and C. Macchiavello, Phys. Rev. A \textbf{83}, 052313 (2011).
\bibitem{Datta08}
A. Datta, A. Shaji, and C. M. Caves. Phys. Rev. Lett. \textbf{100}, 050502 (2008).
\bibitem{Adhi13}
S. Adhikari, and S. Banerjee,  Phys. Rev. A \textbf{86}, 062313 (2012).
\bibitem{Pelli95}
T. Pellizzari, S. A. Gardiner, J. I. Cirac, and P. Zoller, Phys. Rev. Lett. \textbf{75}, 3788 (1995). 
\bibitem{Sho95}
P. W. Shor, Phys. Rev. A \textbf{52}, R2493 (1995).
\bibitem{Elli04}
D. Ellinas, and Ch. Konstandakis, Proc. AIP Conf.  734 (2004).
\bibitem{Shap03}
D. Shapira, S. Mozes, and O. Biham, Phys. Rev. A \textbf{67}, 042301 (2003). 
\bibitem{Norm99}
B. P. Norman, and M. R. Altaba, Phys. Rev. A \textbf{61}, 012301 (1999).
\bibitem{Gaw11}
P. Gawron, J. Klamka, and R. Winiarczyk, Appled Math. and Comp. Sci. \textbf{22}, 493 (2012).
\bibitem{Woo98}
W. K. Wootters, Phys. Rev. Lett. \textbf{80}, 2245 (1998).
\bibitem{Woo01}
W. K. Wootters, Quantum Info. and Comp., \textbf{1} (2001).
\bibitem{NieChu00}
M.A. Nielsen, and I. Chuang, Quantum Computation and Quantum Information (Cambridge University Press, 2002), p 380.
\bibitem{Bruk10}
B. Dakic, V. Vedral, C. Brukner, Phys. Rev. Lett. \textbf{105}, 190502 (2010).
\bibitem{Giro11}
D. Girolami, and G. Adesso, Phys. Rev. A \textbf{83}, 052108 (2011).
\bibitem{Shan13}
S. Chakraborty, A. Kumar, S. Adhikari, in preparation.
\end{thebibliography}
\end{document}